\documentclass[sigconf]{acmart}
\AtBeginDocument{%
  }

\begin{document}

\title{A Measure Based Generalizable Approach to Understandability}

\author{Vikas Kushwaha}
\email{vikask23@cse.iitk.ac.in}
\affiliation{%
  \institution{Indian Institute of Technology Kanpur}
  \country{India}
}

\author{Sruti Srinivasa Ragavan}
\email{srutis@cse.iitk.ac.in}
\affiliation{%
  \institution{Indian Institute of Technology Kanpur}
  \country{India}
}

\author{Subhajit Roy}
\email{subhajit@cse.iitk.ac.in}
\affiliation{%
  \institution{Indian Institute of Technology Kanpur}
  \country{India}
}

\renewcommand{\shortauthors}{Kushwaha et al.}
\newcommand{\todo}[1]{\textcolor{red}{TODO:{#1}}}

\begin{abstract}
Successful agent-human partnerships require that any agent generated information is understandable to the human, and that the human can easily steer the agent towards a goal. Such effective communication requires the agent to develop a finer-level notion of what is understandable to the human. State-of-the-art agents, including LLMs, lack this detailed notion of understandability because they only capture average human sensibilities from the training data, and therefore afford limited steerability (e.g., requiring non-trivial prompt engineering).

In this paper, instead of only relying on data, we argue for developing generalizable, domain-agnostic measures of understandability that can be used as directives for these agents. Existing research on understandability measures is fragmented, we survey various such efforts across domains, and lay a cognitive-science-rooted groundwork for more coherent and domain-agnostic research investigations in future.
\end{abstract}

\settopmatter{printacmref=false}
\setcopyright{none}
\renewcommand\footnotetextcopyrightpermission[1]{}
\pagestyle{plain}

\maketitle

\section{Introduction}
\label{sec:intro}
An agent is an automated system that solves tasks on behalf of its users, or aids them while they solve a problem. For example, consider a robotic agent tasked with obtaining an object in a real world setting. It will explore the environment, develop an information map of its environment relevant to the task, maybe ask its human operator some questions, and then use that information to create a plan to accomplish the task. Consider another example, namely a software agent that assist programmers by highlighting class or function namings in programs that do not align with the logic contained within them, or point out better structuring alternatives for modules and class hierarchies, or provide better code alternatives (e.g., based on newer library versions for example). Such an agent can partner with human programmers in writing more understandable and maintainable code.

Such agent-human partnerships can only work well when the agent communicates in a manner that is understandable to humans \cite{stephanidis2019seven}. Further, open-ended tasks such as creative writing and software design require agents that can be steered according to user's needs. For example, a writer might have their own sensibilities and style about how an essay should be written and structured. In turn, this requires the agent to have a finer-level differentiation over general human cognitive priors and individual user sensibilities, with the ability to expose them as ``steering'' controls to the human user.

However, current agents have poor steerability \cite{chang2024measuring}. For example, in one experiment, an LLM-based agent defaulted to a pleasant ``American'' tone, even when prompted to change their tone to increased sadness \cite{chang2024measuring}. Similarly, a software designer may have certain expectations about code structuring, class hierarchy organization or variable naming policy, and while a programming-assist agent might produce reasonably understandable code, it might not be able to follow the policies that designer intends without a model of how humans work with hierarchies, analogies and saliency relationships. Likewise, an author of a technical specification may be fine with sacrificing some understandability at the cost of being thorough, but if a neural agent has not seen enough of such contexts during training, it could still produce text that would be in common writing style, inadvertently sacrificing the thoroughness for understanding. 

We think such agents lack steerability \cite{chang2024measuring} because of their over reliance on data. Whereas current agents are good at capturing average human priors and sensibilities indirectly, since they are trained on human-generated data, they can not differentiate between various aspects of human sensibilities at a finer level as we saw in the examples above. In puzzles involving visual objects, patterns and numeric elements \cite{chollet2024arc}, we see that LLMs are often unable to generate or follow strategies for solving previously unseen visual reasoning problems, even though these puzzles rely on (and are generalizations of) ``human core knowledge'' prior \cite{spelke2007core} about such elements.

One solution for solving such problems is to generate synthetic pre-training data using cognitive models as shown in works such as by Bourgin et al. \cite{bourgin2019cognitive}, but this approach is specific to each use case (e.g., various kinds of visual, textual, math, decision-making scenarios). This approach might work for limited use cases, but elsewhere can be tedious and potentially leading to the ``averaging out'' when multiple aspects are involved.

We propose an alternative, namely that agents reduce their reliance on training data for such purposes and instead take advantage of research in human cognition (e.g., cognitive architectures \cite{laird2017standard}, models of text comprehension \cite{mcnamara2009toward}, motivation \cite{schmidhuber2010formal, oudeyer2007intrinsic} and usability \cite{baecker2014readings}) to develop a more general solution. By modeling alignment with human sensibilities as optimization directives for these agents, we could develop agents capable of more explicit and finer differentiation over aspects such as comprehensibility, compared to what could be indirectly captured from noisy human data during training. But, to specify ``alignment with human sensibilities'', we need measures of these sensibilities for the information entities (e.g., text, program code, user interfaces) produced by the agents. 

Once again, we could adopt artifact dependent domain-specific measures for each of the human priors and sensibilities (e.g., motivation, understandability, emotional experience) paired with various human-agent communication artifacts (e.g., text, code, visualizations). In fact, there exist several such measures -- text and code complexity metrics \cite{wang2012formal, mccabe1976complexity, halstead1977elements}, video perceptual quality measures \cite{van1996perceptual}, code cohesion measures \cite{morasca1997towards, nicolaescu2015evolution}, inconsistency measures for logic systems \cite{hunter2005approaches, grant2006measuring}, to name a few. But, since these measures can be traced back to human abilities and limitations such as in motivation, perception, attention, memory, or information processing, we posit that there must exist a set of more fundamental ``abstract'' measures rooted in cognitive psychology. The measures listed above for text, code or video are simply domain-specific operationalizations of these underlying abstract cognitive-psychology measures.


As a starting point, this paper focuses on one kind of human priors and sensibilities, namely understandability. We propose a cognitive-science-rooted framework to organize fundamental aspects of understandability into six coherent dimensions: \textit{perceptual quality}, \textit{memory cost}, \textit{pattern decodability}, \textit{cohesion}, \textit{logical consistency} and \textit{semantic fit}. While existing research on understandability measures is fragmented, we survey various such efforts across domains from the perspective of this framework. This should highlight research gaps and motivate future conversations for developing more generalized measures that can work ``as a whole'' across a variety of information entities (text, code, etc.).


\section{Background and related work}
\label{sec:process}

\subsection{Cognitive modeling}
The field of cognitive science concerns with understanding processes in the human mind, examples are attention, speech, comprehension and learning. One way in which psychologists study these phenomena is by: 1) developing specific hypotheses about the process under study, 2) developing computational models and architectures that implement these hypotheses and 3) evaluating the predictions of the models against actual human behavior for a specific task. Some examples are \cite{kintsch2013construction, von1995program, oudeyer2007intrinsic, mervis1981categorization}. A longstanding criticism of task specific cognitive models has been that they lead to a fragmented approach and make cumulative progress difficult \cite{laird2017standard}. As a solution, calls have been made for integrated models of cognition. Cognitive architectures (e.g., ACT-R, Soar, Sigma) \cite{laird2017standard} are a step in fulfilling that goal. ``However, there has historically been little agreement either across or within specialties as to the overall nature and shape of this architecture'' \cite{laird2017standard}. The larger goal of unified theories of cognition \cite{newell1994unified} remains a work in progress \cite{laird2017standard}. 

\subsection{Cognitive models of understanding}

Understanding is a specific kind of cognitive process and similar issues befall it. While there is a general idea that 1) understanding involves some sort of fit between data or the artifact being understood with the frame or schema (e.g., a mental structure for organizing and perceiving new information) \cite{newell1994unified, russell1993cost, naumer2008sense} and 2) that it involves activation and evaluation of relevant information in memory \cite{marinier2006cognitive}, more detailed models of comprehension are fragmented and domain-specific. For example, there are models of text comprehension \cite{mcnamara2009toward}, program comprehension \cite{von1995program}, language comprehension \cite{johnson1981comprehension} and so on. However, while the details of comprehension models may vary both within and across domains, the key aspects are largely shared. With this in mind, we make an attempt to describe major aspects of the human process of understanding, based on emerging ideas. Our main aim here is to figure out the key aspects of the underlying process of understanding and delineating what aspects of an artifact (e.g., text, code) contribute to their understandability. This gives us an initial base of understandability aspects to be investigated and improved upon in future. 

\begin{figure*}[h]
  \centering
  \includegraphics[width=0.75\linewidth]{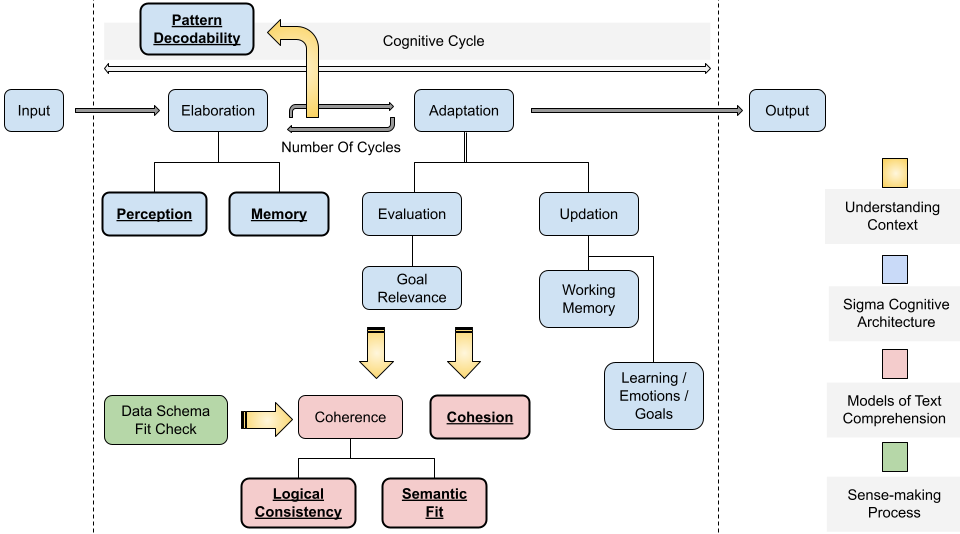}
  \caption{Emerging landscape across models of understanding process}
  \label{fig:figure1}
\end{figure*}

Although cognitive architectures do not cover every detailed aspect of cognition, they are the most accepted models of cognitive processes. The emerging consensus among such architectures is that the human cognitive processes involve perception, memory and a cognitive cycle \cite{laird2017standard}. For example, the recent Sigma architecture \cite{rosenbloom2016sigma} draws lessons from previous architectures, and posits that the human cognitive process is mainly a two phased cyclic process (See Figure~\ref{fig:figure1}). The first phase (elaboration) deals with perceiving the information at hand, retrieving related information from memory and reasoning to draw conclusions from it. The second phase (adaptation) is largely about updating the working memory state and long term memory (such as in learning), and works based on evaluating how far the current state is from the person's desired goals. While memory and perception are important across all kinds of cognitive processes (e.g., recognizing objects, solving puzzles), such evaluations play an essential role in the process of comprehension \cite{marinier2006cognitive}.

Whereas mechanisms involved in perception, memory (retrieval and updation) are largely task independent, evaluations can be more task dependent (e.g., novelty of stimulus for learning and motivation, matching with goal state for simple planning). For comprehension, models across domains (text comprehension \cite{kintsch2013construction}, language comprehension \cite{johnson1981comprehension}, code comprehension \cite{von1995program} and general sense-making models \cite{russell1993cost, naumer2008sense}) largely suggest two kinds of evaluations, namely cohesion and coherence (also referred to as data-schema fit \cite{mekler2019framework}). Cohesion is about making sense of a structure through its individual constituents (e.g., words in a sentence, methods in a program class), while coherence is about relation between multiple structures (e.g., how paragraphs make an argument, logical arrangement of UI screens, relations of a program class with other classes). In terms of measures, cohesion is often characterized as relatedness or similarity between the constituents of a structure (e.g., use of the same word, pronouns) while coherence can be seen as a constraint satisfaction process involving varieties of constraints based on the kinds of elements and relationships involved \cite{thagard2002coherence}. Coherence constraints can be of broadly two main types: 1) logical consistency between declaration or proposition like elements and 2) semantic compatibility or fit between elements in relationships such as analogies, explanations, interpretations and conceptual associations \cite{thagard2002coherence}. Finally, the number of cognitive cycles is also related to the inherent complexity of the artifact being understood \cite{madl2011timing}; we can operationalize it as some measure of difficulty of mining useful patterns present in an artifact (or ``pattern decodability'').




\section{Organizing understandability into dimensions}
\label{sec:aspects}

Summarizing the emerging landscape across cognitive models of understanding, we propose six dimensions of understandability: \textit{perceptual quality}, \textit{memory cost}, \textit{pattern decodability}, \textit{cohesion}, \textit{logical consistency} and \textit{semantic fit}. We highlight these dimensions with underlines in Figure ~\ref{fig:figure1} that summarizes the cognitive processes of understanding, synthesized across various recent models (namely, cognitive architectures, models of comprehension and sensemaking). The rooting of these models in cognitive architectures allows for coherence, while the use of abstractions from cognitive psychology allows for generalizability across domains.  


In this section, we discuss each of these dimensions in depth, summarize the existing research on understandability measures from various domains (e.g., software engineering, theoretical computer science, psychology) as operationalizations of these dimensions, and highlight a path forward into generalizing and adopting them for newer domains.



\subsection{Perceptual quality}
\label{sec:processing}
Humans have innate tendencies in what kinds of information we privilege in processing over others, based on the kind of environment we evolved in. We observe these tendencies in human perception. For example, as per Gestalt laws of grouping \cite{wertheimer2012investigations}, we perceive patterns in stimulus based on certain principles such as proximity, similarity, continuity, closure and connectedness. We can also see similar examples in other fields, such as how we process variable names in software programs or focus/miss some elements on a user interface \cite{oviatt2000perceptual}. 

Such observed differences can be studied as relationships between concreteness, perceptual experiences and processing speed. Based on a baseline of human ranking of concreteness in words \cite{paivio1968concreteness}, Walker et al. \cite{walker1999concrete} found that concrete words are easier to recall than abstract words. Likewise, interplay of concreteness and precision has been found to play a role in cognitive processing efforts \cite{iliev2017paradox} \cite{frank2010uncertainty}, even though the effect of precision may partly be explained based on larger memory requirements it entails. Connell et al. \cite{connell2012strength}, in another experiment, found that strength of perceptual experience is a better predictor of word processing performance than concreteness or imageability. We also see perceptual quality measures characterizing entities such as user interfaces \cite{hsiao2006gestalt} and videos \cite{van1996perceptual} based on models of human perception.

By characterizing these qualities, we can understand what kind of information is processed faster by humans due to better encoding or specialized processing for them. Creating more general measures may involve combining perceptual measures across different kinds of sensory information and richer perceptual information datasets for concepts.

\subsection{Memory cost}
\label{sec:memoryCost}
Understanding process requires obtaining the concepts and relationships which are part of, or related to, the information entity. Memory plays a pivotal role in recognition and recall of information, and holding the state of this process. Human memory has been broadly classified into working (short term) memory and long term memory \cite{baddeley2013essentials}. Memory cost imposed by an information entity depends both on the number of elements or concepts involved, as well as the amount and type of connections in their relationship graphs. We can see the role of memory cost in influencing usability heuristics such as recognition over recall, use of conventions and consistency in design \cite{nielsen1994usability}.

In software engineering, there have been multiple attempts at defining program complexity metrics in order to measure program's readability. Largely, most such metrics deliberately or otherwise focus on the memory cost of the program for the programmer trying to comprehend the program. We will list four examples here.

One of the earlier such metrics is Cyclomatic complexity \cite{mccabe1976complexity}. It is a measure of the number of linearly independent paths in a program. More the number of such paths, more difficult it is to test the program or comprehend all possible scenarios, as it would require more time and working memory to keep track of each scenario. With Halstead measures \cite{halstead1977elements} the goal is to define code measures related to program size and their relationship with the effort required to understand the program. A code with a larger number of distinct operations and dependencies carried through operands would be tougher to understand. This aligns with our understanding of the need of working memory to hold states across the understanding process. Long term memory is needed to fetch the information related to those operations. Spatial measures \cite{douce1999spatial} are based on the distance between components of a program, such as distance between definition and usage of a function. Larger the distance, more efforts during comprehension are expended either to locate the other components or to keep them in memory to avoid locating them. Cognitive complexity \cite{campbell2018cognitive} measures the amount of flow breaks and nestings in the code, as more of these mean more state needs to be tracked by the reader of the code.

Apart from programs, we also see works \cite{wang2012formal} on characterizing complexity of text based on sentence structure. Complex sentence structure requires the reader to hold more state to comprehend the text.

In a generalized setting, we can see information entities in terms of conceptualizable elements such as symbols, words, visual objects, etc and their relationships. Number and familiarity of these elements can help in measuring the long term memory needs of the entity. Any subset of these elements, which are highly related with each other would need to be held in working memory together to derive conclusions or fetch associated information. This can be used to characterize the working memory requirements of an information entity. It is worth looking at memory models in cognitive architecture such as ACT-R \cite{anderson2004integrated} for ideas on developing measures for memory cost of information entities.

\subsection{Pattern decodability}
\label{sec:patternDec}
Information entities contain cues which the understander needs to look up in memory to get to the associated concept. Sometimes the entity contains patterns which need to be recognized and decoded before they can be looked up in memory or used in further processing. While some patterns are straightforward to recognize, there are information entities that need more cognitive cycles before the embedded patterns become obvious. Consider the example of code refactoring in software engineering. It requires the developer to detect patterns across the program to see if they can be pulled into a shared module, or written more succinctly using a different construct or abstraction.

When it comes to defining measures for difficulty of recognizing patterns, there have been attempts to define it based on compressibility of strings. In computational information theory, Kolmogorov complexity \cite{li2008introduction} gives us a measure for lack of patterns in a string. However, it is not always computable \cite{li2008introduction}. Measures related to computable approximations of Kolmogorov complexity such as Lempel Ziv complexity \cite{lempel1976complexity} can be used to measure the diversity of patterns in data. It has seen practical applications and has been used to study the pattern diversity of neurological data, DNA sequences, and other kinds of multi dimensional data \cite{zozor2005lempel}.    

Notion of computational depth \cite{antunes2006computational} adds on top of above measures characterizing the existence of patterns. It measures the amount of non-trivial patterns in a string as a difference between time restricted Kolmogorov complexity and its traditional counterpart \cite{antunes2006computational}. A string having a short compressed version but requiring considerable computation time to uncompress will have large computational depth, compared to random strings or strings with trivial patterns. Further, there are computable variants of computation depth such as Lempel-Ziv depth \cite{jordon2023pushdown}. If we can represent any information entity as a binary string, the above collection of measures can serve us in quantifying the amount of useful patterns in the entity and how computationally intensive mining those patterns is.

\subsection{Cohesion}
\label{sec:cohesion}
Clear boundaries between concepts make the concepts less confusing. More tightly linked the constituents of a concept or structure, the more cohesive it is. Real world often has fuzzy boundaries and interconnected ideas. However, organizing formations play an important role in allowing processing of information as easier to deal with units. Hence, the quality of these formations plays a key role in improving understandability. We can see this in software design when trying to modularize and structure the code, in text, when ensuring the cohesion between neighboring sentences, or in visualizations, when trying to illustrate related concepts (e.g., via a flowchart, block diagram or taxonomy tree).

This has been modeled and measured in software engineering and especially object oriented programming in terms of coupling and cohesion \cite{morasca1997towards, nicolaescu2015evolution}. Cohesion is a measure of the degree to which elements inside a module belong together. Coupling is a measure of how closely connected two modules are. There are many variants of coupling/cohesion such as logical, temporal, semantic \cite{marcus2008using} etc. Related components in a module should have high internal cohesion, but lower coupling with other less related components outside the module. We also see similar ideas in research on categorization \cite{gennari1989models} of instances of a concept. We split the category if cluster of their instances have a high variance and the split clusters are not tightly coupled, and combine them otherwise. Models of text comprehension have proposed argument overlap between sentences as a measure of cohesiveness \cite{mcnamara2009toward}.

Given a graph based representation for a domain, measures such as coupling and cohesion measures can be generalized to characterize the cohesiveness for various kinds of information entities. When dealing with instances of a concept with property values drawn from a continuous scale, clustering based cohesion characterization can be used. We can also attempt combining the two approaches into one general approach.

\subsection{Logical consistency}
\label{sec:logicalConsistency}
Information entities in real life often contain inconsistencies between their declaration or proposition like elements. Humans are able to reason under uncertainty and inconsistent data to a certain extent, but in general, the greater the inconsistency, the lesser the entity makes sense to the reader, and requires more processing (and thus, greater time and cognitive effort) before reasoning can be performed on it. Measures of inconsistency can also be used to evaluate how far a given entity is from satisfying a given set of constraints or goals. These constraints/goals can come from the problem which the human is trying to solve, or from the previously learned schema used for making sense of the data.

The field of approximate logic allows us to model logic and reasoning under inconsistency. Various frameworks at the intersection of AI and logic have been proposed to make inconsistent reasoning tractable \cite{lozinskii1994resolving, darwiche1997logic}. Further, inconsistency measures have been proposed to characterize the amount of inconsistency present in the information entity \cite{hunter2005approaches, grant2006measuring}. These measures depend on alternative logic systems apart from classical logic to accommodate inconsistent reasoning. 

Future research in this area can take up both the problem of defining better inconsistent logic systems, which can express the real world data better, as well as compute the inconsistency measure efficiently, especially when dealing with large numbers of predicates.

\subsection{Semantic fit}
\label{sec:semanticFit}
A concept should satisfy certain optimality principles as to how the concept is semantically organized with respect to other concepts. We will give three cases here to illustrate the idea. (1) We tend to prefer conceptualizations which are between most general and most specific \cite{bobick1987natural, mervis1981categorization} in general usage. It can be seen as a tendency to balance between memory cost and predictive power. So, when referring to a specific type of chair such as ``Windsor'' in day to day usage context, we prefer to use term like ``Chair'' over more general term like ``Furniture'' and more specific term like ``Windsor''. In more specific contexts, for example a catalog of chairs, we would use the specific terms as conveying that extra information is required there. (2) Concepts representing analogies, metaphors, summaries, counterfactuals, etc should capture the saliency of input conceptual spaces from which they are derived. Fauconnier et al. \cite{fauconnier1998conceptual} talk about such optimality principles for conceptual blends such as analogies and counterfactuals. (3) Abstraction jumps should be avoided when organizing concepts in a structure such as a concept hierarchy. Consider a concept hierarchy of countries with the ``World'' as root node, continents at second levels and individual nations at third. Now, if we place a country ``Xyz'' at second level alongside continents, it would be surprising for the reader and would not make sense to them. Concepts which satisfy such optimality principles are easier to understand.

If we look at domains such as software engineering, we would find examples of semantic fit, starting from naming of classes and functions. We prefer not too long names that can capture the saliency of what we are trying to do in the code. Class hierarchies, configuration or API specifications should avoid abstraction jumps within the same level, so that they do not surprise the developer. Similar examples can be seen with names and menu structures in user interfaces. 

However, despite its importance, we did not find much prior research in this area that we can capitalize on in further formalizing the idea, or creating measures for semantic fitness. In other words, this understandability aspect is open for future research contributions.

\section{Concluding remarks and future directions}
\label{sec:discussion}
The notion of understandability has been operationalized in various domains in fragmented ways, with several measures and rules of thumb existing to make content more understandable to their readers (e.g., usability heuristics for UI design, metrics for clean code, text comprehensibility measures used in psychology experiments). This fragmentation not only hurts building a coherent body of knowledge in HCI, but is also a practical limitation, such as when building agents that can be steered by humans in generating understandable content for various tasks. 

However, such efforts in various domains are meant to deal with fundamental human cognitive abilities and limitations, and thus in this paper, we aimed to derive six dimensions of understandability from widely accepted cognitive models of how humans go about understanding ``stuff''. Our six dimensions, namely \textit{perceptual quality}, \textit{memory cost}, \textit{pattern decodability}, \textit{cohesion}, \textit{logical consistency}, and \textit{semantic fit}, each map on to various stages of the cognitive processes underlying understanding, and the existing measures of understandability map on to these dimensions. 

Apparently, our six dimensions of understandability have properties desirable in a good theory \cite{sjoberg2008building}: \textit{parsimonious} -to the extent the grounding in cognitive psychology let us be-, have \textit{explanatory} and \textit{predictive} power (as to why an artifact can be hard to understand and how to fix it), \textit{generalizability}, by way of using cognitive science abstractions not specific to any application domain, and \textit{practical applicability} such as when measures are developed and used in specific domains (as described earlier). 

Our effort is also complementary to existing theories that underlie usability. For example, the information foraging and sense-making theories \cite{pirolli1995information, russell1993cost, naumer2008sense} have been influential by simply framing human information seeking as a value-cost optimization. These works show that evaluating an environment's cost value offerings can highlight issues of usability. Our six dimensions of understandability can provide a framework in which to optimize the costs of consuming information, once a forager has foraged for it. Likewise, it also serves as a framework to more concretely model sense-making costs involved in an environment or information artifact.

However, despite solid theoretical grounding, we point out that our effort is preliminary, and, as with any theory-building effort, our six dimensions need validation. Thus, an essential, immediate future work is to critically evaluate the completeness, appropriateness and generalizability of our dimensions via user studies. In turn, this will also serve as indirect validation of the cognitive models in terms of their generalizability and completeness. 

Another immediate line of future possible work would be to explore each of these understandability dimensions on a finer level. We see that semantic fit represents a relatively less explored area when it comes to developing formalizations and measurable characterizations. It would benefit from more research focus. That said, each of the individual dimensions present avenues for developing better measures for those aspects.

More tactically, this paper opens up several avenues for research in the near future. One set of problems relate to the implementation of agents that would utilize these six dimensions of understandability and their measures. For example, what kind of optimization algorithms can be used with directives based on these dimensions of understandability? What is the internal representation that the agent should use for these information entities (i.e., code, text, user interfaces)? The representation should be generalizable and composable to accommodate creation and updation of a variety of such entities while making it easier to compute understandability measures on it. Another set of problems worth exploring pertains to the role these characterizations can play in real world problem solving. Open world planning remains a largely unsolved problem in AI \cite{kanervisto2022minerl}. Would an agent capable of characterizing understandability and using them as optimization directives perform better on these problems, as one would speculate based on connections between Polya's model of problem solving and AI planning \cite{newell1983heuristic}. Likewise, do understandability directed agents produce more maintainable software code in terms of number of bugs or time taken to implement new features?


Finally, going beyond understandability, this paper offers preliminary insights into a method for drawing from theories of cognitive psychology to make them practical to HCI research. For example, there are multiple kinds of human priors and sensibilities that are relevant to HCI (e.g., user motivation \cite{schmidhuber2010formal, oudeyer2007intrinsic}, exploration tendencies \cite{pirolli1995information}). They deserve research conversations of their own, but can take a similar approach to us in organizing and exploring those aspects with a solid grounding in theories.




\bibliographystyle{acm}
\bibliography{paper}



\end{document}